\documentclass[12pt,preprint]{aastex}

\newcommand{\etal }{{et al.} }
\newcommand{\msun}{\thinspace M_\odot}

\def\lesssim{\mathrel{\hbox{\rlap{\hbox{\lower4pt\hbox{$\sim$}}}\hbox{$<$}}}}
\def\gtrsim{\mathrel{\hbox{\rlap{\hbox{\lower4pt\hbox{$\sim$}}}\hbox{$>$}}}}
\newcommand{\cm}{\,{\rm cm}^{-3} } 
\newcommand{\km}{\,{\rm km\, s}^{-1}} 
\newcommand{\nc}{n_{\rm c} } 
 
\newcommand{\mj}{M_{\rm Jup}}
\newcommand{\rcri}{R_{\rm c} }
\newcommand{\mdot}{M_\odot\,{\rm yr}^{-1} }
\newcommand{\tc}{t_{\rm c}}

\shorttitle{Brown Dwarf Formation}
\shortauthors{Machida \etal 2009}

\begin{document}

\title{First Direct Simulation of Brown Dwarf Formation in a Compact Cloud Core}

\author{Masahiro N. Machida\altaffilmark{1},  Shu-ichiro Inutsuka\altaffilmark{1,2}, and Tomoaki Matsumoto\altaffilmark{3}} 

\altaffiltext{1}{Department of Physics, Graduate School of Science, Kyoto University, Sakyo-ku, Kyoto 606-8502, Japan; machidam@scphys.kyoto-u.ac.jp}
\altaffiltext{2}{Department of Physics Nagoya University Furo-cho, Chikusa-ku Nagoya, Aichi 464-8602; inutsuka@tap.scphys.kyoto-u.ac.jp}
\altaffiltext{3}{Faculty of Humanity and Environment, Hosei University, Fujimi, Chiyoda-ku, Tokyo 102-8160, Japan; matsu@i.hosei.ac.jp}

\begin{abstract}
Brown dwarf formation and star formation efficiency are studied using a nested grid simulation that covers five orders of magnitude in spatial scale ($10^4$--$0.1$\,AU).
Starting with a rotating magnetized compact cloud with a mass of $0.22\msun$, we follow the cloud evolution until the end of main accretion phase.
Outflow of $\sim5\km$ emerges $\sim100$\,yr before the protostar formation and does not disappear until the end of the calculation.
The mass accretion rate declines from $\sim10^{-6}\mdot$ to $\sim10^{-8}$--$10^{-12}\mdot$ in a short time ($\sim10^4\,$yr) after the protostar formation.
This is because (1) a large fraction of mass is ejected from the host cloud by the protostellar outflow and (2) the gas escapes from the host cloud by the thermal pressure.
At the end of the calculation, $74\%$ ($167\mj$) of the total mass ($225\mj$) is outflowing from the protostar, in which $34\%$ ($77\mj$) of the total mass is ejected by the protostellar outflow with supersonic velocity and $40\%$ ($90\mj$) escapes with subsonic velocity. 
On the other hand, $20\%$ ($45\mj$) is converted into the protostar and $6\%$ ($13\mj$) remains as the circumstellar disk.
Thus, the star formation efficiency is $\epsilon = 0.2$.
The resultant protostellar mass is in the mass range of brown dwarfs.
Our results indicate that brown dwarfs can be formed in compact cores in the same manner as hydrogen-burning stars, and the magnetic field and protostellar outflow are essential in determining the star formation efficiency and stellar mass.
\end{abstract}
\keywords{ISM: clouds---ISM: jets and outflows---ISM: magnetic fields ---MHD---stars: formation---stars: low-mass, brown dwarfs}

\section{Introduction}
There are several scenarios for the formation of brown dwarfs \citep[see,][]{whitworth07}, though many observations indicate that brown dwarfs are formed through the gravitational collapse of a molecular cloud core \citep[see,][]{luhman07}.
Observations around young brown dwarfs show the existence of circumstellar disks \citep[e.g.,][]{pascucci03, luhman05} and of gas accretion onto the young brown dwarfs \citep[e.g.,][]{white03}.
Recently, \citet{whelan05} observed an optical jet around a young brown dwarf of $\rho$ Oph 102.
In addition, \citet{phan08} observed molecular outflows from the same object.
To date, several outflows have been observed around young brown dwarfs \citep{whelan09a, whelan09b}.
Since outflow is typical in the star formation process, it is considered that these flows are direct proof that brown dwarf formation occurs through the gravitational collapse of a molecular cloud core. On the other hand, observations indicate that the prestellar core mass function (CMF) resembles the initial mass function \citep[IMF, e.g.,][]{motte98}, suggesting that the effective reservoirs of mass required for the formation of individual stars including brown dwarfs are already determined at the prestellar core stage \citep{andre08}.
The similarity of CMF and IMF implies that the prestellar cores are on a one-to-one correspndence with protostars, in which the star formation efficiency is expected to be $\epsilon \equiv M_{\rm star}/M_{\rm core}\sim20$--$50$\%.

Not all the mass of a prestellar core can convert into a star, because the outflow from a protostar can remove a significant mass from the natal cloud core \citep {whitworth07}.
\citet{matzner00} has pointed out that since the gas accretion onto a protostar is impeded by the protostellar outflow, only a fraction of the mass can convert into the protostar, with the remainder being blown out.
They modeled the prestellar outflow and analytically estimated a star formation efficiency of $\epsilon \simeq$ 25--70\%.

We can presume that brown dwarfs are born in more compact (or less massive) cloud cores, because they are less massive than stars.
Taking a star formation efficiency of $\epsilon = 30$\%, brown dwarfs with masses of $13$--$75\,\mj$, where $\mj$ is the Jovian mass, can be formed in cores with masses of $43$--$250\,\mj$.
Cloud cores with such masses have been observed in several surveys \citep[e.g.,][]{motte01a,sandel01,onishi02}.
In this paper, we study the evolution of a compact molecular cloud core and the formation of a brown dwarf.

\section{Model and Numerical Method}
To study the evolution of star-forming cores in a large dynamic range of spatial scale, a three-dimensional nested grid method is used, in which the equations of resistive magnetohydrodynamics with a barotropic equation of state are solved (see Eq. [1]--[3] and [5] of \citealt{machida07}).

In the collapsing cloud core, we assume protostar (or proto-brown dwarf) formation occurs when the number density exceeds $n > 10^{14}\cm$ at the cloud center.
To model the protostar, we adopt a sink around the center of the computational domain.
In the region $r < r_{\rm sink} = 0.5\,$AU, gas having a number density of $n > 10^{14}\cm$ is removed from the computational domain and added to the protostar as a gravity in each timestep.
In addition, inside the sink, the magnetic flux is removed by the Ohmic dissipation (The detailed description will appear in a subsequent paper).

As the initial state we take a spherical cloud with a Bonnor--Ebert (BE) density profile that extends up to twice the critical BE radius ($r = 2\, \rcri $).
Outside the sphere ($r > 2\, \rcri$), a uniform density is adopted.
For the BE density profile, we adopt a central density of $\nc = 5\times 10^{7}\cm$ and an isothermal temperature of $T=10$\,K. 
For these parameters, the critical BE radius is $\rcri = 650$\,AU.
Thus, the radius of the initial sphere is $r = 2\, \rcri = 1300$\,AU.
To promote the contraction, we increase the density by a factor of 1.68.
The density contrast between the cloud center and ambient medium is about 80.
The mass within $r < 2\, \rcri$ ($r < \rcri$) is $0.22\msun$ ($0.11\msun$) corresponding to $225\mj$ ($112\mj$).
In this paper, we call the sphere with $r<2\,\rcri$ the host cloud.
The cloud rotates rigidly with $\Omega_0 = 2.5 \times 10^{-12}$\,s$^{-1}$ around the $z$-axis in the region of $r<2\, \rcri$, while a uniform magnetic field ($B_0 =212 $ $\mu$G) parallel to the $z$-axis (or rotation axis) is adopted over the whole computational domain.
In the region of $r<2\, \rcri$ ($r < \rcri$), the ratios of thermal $\alpha_0$, rotational $\beta_0$, and magnetic $\gamma_0$ to the gravitational energy are $\alpha=0.42$ (0.5), $\beta=0.04$ (0.016), and $\gamma=0.13$ (0.04), respectively.
The gravitational force is ignored outside the host cloud ($r>2\,\rcri$) to mimic a stationary interstellar medium.

To calculate over a large spatial scale, the nested grid method is adopted \citep[for details see ][]{machida05a,machida05b}. 
Each level of a rectangular grid has the same number of cells ($ 64 \times 64 \times 32 $).
The calculation is first performed with five grid levels ($l=1$--$5$).
The box size of the coarsest grid $l=1$ is chosen to be $2^5 \rcri$.
Thus, a grid of $l=1$ has a box size of $\sim 2\times 10^4$\,AU.
A new finer grid is generated before the Jeans condition is violated.
The maximum level of grids is restricted to $l_{\rm max} \leqq 10$.
The $l=10$ grid has a box size of 40\,AU and cell width of 0.63\,AU.

\section{Results}
Since we adopted an unstable core with a centrally peaked density profile as the initial state, the central part of the cloud collapses first and the gas density increases with time.
In the collapsing cloud core, just after the central number density reaches $\nc \simeq 10^{11}\cm$, the first adiabatic core \citep[hereafter the first core,][]{larson69,masunaga00} is formed with shock $8.35\times10^3$\,yr after the cloud collapse begins.
The first core has a disk-like shape at its formation, with a size of 18\,AU in the cylindrical radial direction and 3.2\,AU in the vertical direction with a mass of $1.1\times 10^{-2}\msun$.
A low-velocity outflow (hereafter, outflow) of $\sim2\km$ appears around the center of the collapsing cloud 121\,yr after the first core formation.
This kind of outflow in a collapsing cloud has been reported in many other studies \citep[e.g.,][]{tomisaka02,banerjee06,hennebelle08}.
The central density exceeds $\nc > 10^{14}\cm$ and the protostar (hereafter, the proto-brown dwarf) is formed 242\,yr after the first core formation.
Thus, the outflow begins to be driven $\sim100$\,yr {\em before} the proto-brown dwarf formation.
We calculated the cloud evolution $2.3\times 10^4$\,yr after the proto-brown dwarf formation, in which the first core increases in size keeping a disk-like structure and smoothly becomes the circumstellar disk (hereafter the circum-brown dwarf disk) with time.
The outflow continues to be driven from the circum-brown dwarf disk and extends up to $\sim5\times 10^3$\,AU with a maximum speed of $7.2\km$.
From the start of the cloud collapse we calculated the cloud evolution for $2.9\times10^4$\,yr, corresponding to 11.7\,$t_{\rm ff,0}$, where $t_{\rm ff,0}$ is the freefall timescale at the center of the initial host cloud.

Figure~\ref{fig:1} shows the mass accretion rate for the proto-brown dwarf, and the mass of the proto-brown dwarf, outflowing gas, and circum-brown dwarf disk against time ($\tc$) after the proto-brown dwarf formation.
The outflowing gas is defined as gas with velocity $v_r > c_{\rm s}$ for the whole computational domain, while the disk is defined as gas with $v_r < c_{\rm s}$ and $\nc > 5\times 10^{12}\cm$ inside the host cloud.
Figure~\ref{fig:1} shows that the mass accretion rate remains almost constant at $\dot{M}\sim 10^{-6}\msun$\,yr$^{-1}$ for $\tc < 1.5\times 10^4$\,yr, while it suddenly drops and has a very small value of $\dot{M} \sim 10^{-8}$--$10^{-12}\mdot$ for $\tc > 1.5\times 10^4$\,yr.
Note that it slightly decreases with time even for $\tc > 1.5\times 10^4$\,yr.
Reflecting the accretion rate, the mass of the proto-brown dwarf continues to increase for $\tc < 1.5\times 10^4$\,yr, while it remains constant for $\tc > 1.5 \times 10^4$\,yr.
At the end of the calculation, the mass of the proto-brown dwarf is $M_{\rm ps} = 45\mj$.
Since the mass accretion rate is very small for $\tc > 1.5 \times 10^4$\,yr, this object cannot acquire enough mass to become a hydrogen-burning star and hence evolves into a brown dwarf.
On the other hand, the disk mass continues to increase for $\tc < 1.9\times 10^4$\,yr and saturates at $M_{\rm disk}\sim10\mj$.
Finally, the disk mass reaches $M_{\rm disk}=13\mj$.
Even if all the mass of the disk falls into the central object, the central object is within the brown-dwarf mass range ($M_{\rm ps} + M_{\rm disk} = 58\mj$).

\begin{figure}
\includegraphics[width=140mm]{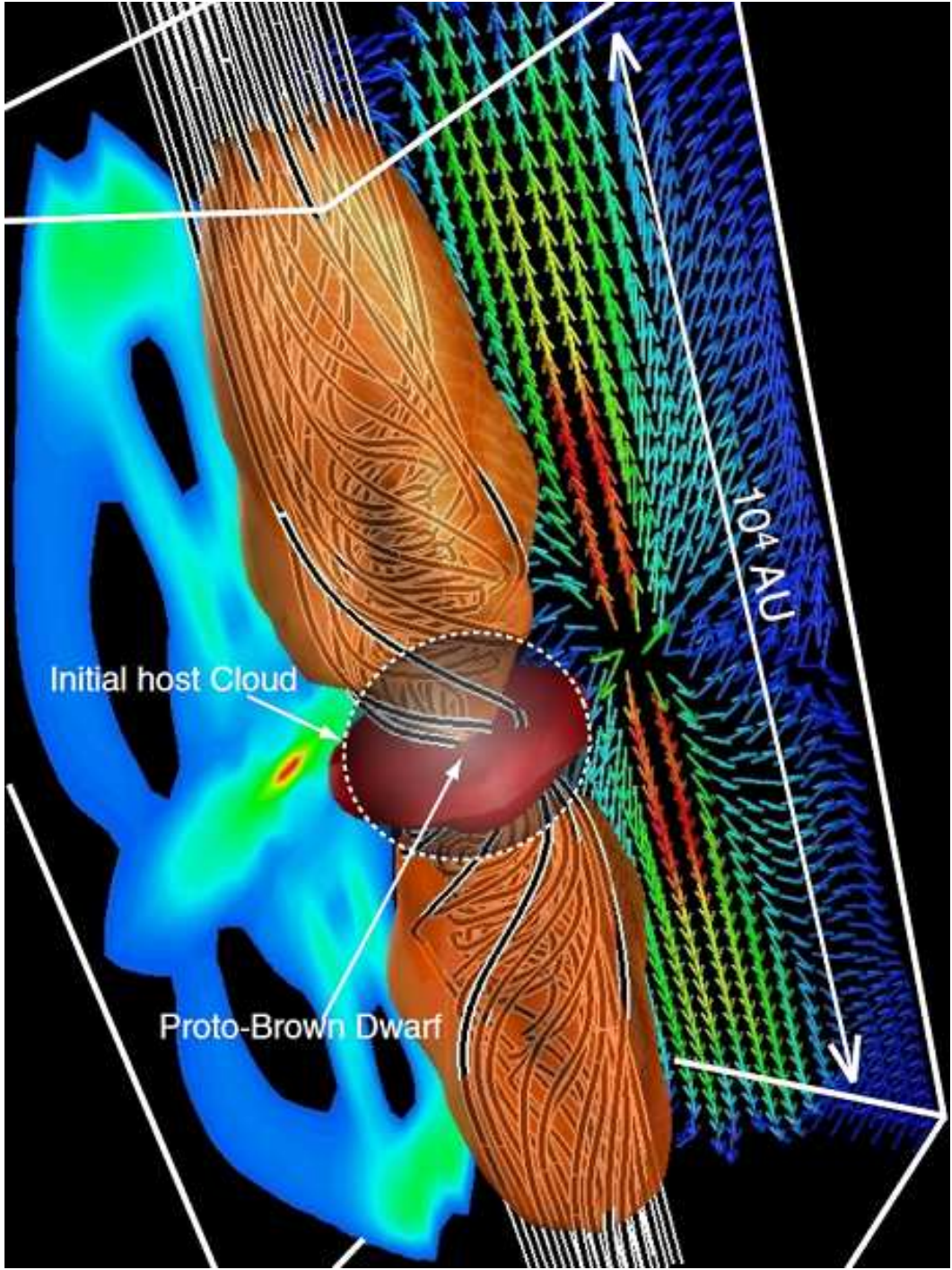}
\caption{
Three dimensionsal structure in the second largest grid ($l=2$).
The structure of the outflow ($v_z > c_s$) is shown by the transparent red surface.
The structure of the high-density region ($ n = 2 \times 10^6\cm $; {\it red-isodensity surface}) and magnetic field lines ({\it black and white streamlines}) are plotted.
The density contours ({\it colors}) on the $x=0$ plane and velocity vectors ({\it arrows}) on the $y=0$ plane are projected on each wall surface.
The white dotted line and transparent black sphere show the initial host cloud of $r = 2 \rcri$.
}
\label{fig:2}
\end{figure}

Figure~\ref{fig:1} shows that the mass of outflowing gas increases over time and reaches $M_{\rm out}=128\mj$.
Although outflow weakens for $\tc > 2\times 10^4$\,yr, it continues to be driven from the disk until the end of the calculation.
Figure~\ref{fig:2} shows the configuration of the outflow.
The iso-velocity of $v_z = c_{\rm s}$ is represented by the transparent red surface, inside which the gas is outflowing from the center of the cloud with supersonic velocity.
The red surface at the center indicates an iso-density of $n=2\times10^{6}\cm$, which almost corresponds to the density at the border between the initial host cloud and the envelope ($n_{\rm amp}=1.1\times 10^6\cm$).
The sphere enclosed by the white dotted circle ($r=2r_{\rm cri}$) represents the initial host cloud. 
The figure shows that the outflow penetrates the host cloud to reach $\sim5\times 10^3$\,AU, about four times larger than the initial host cloud ($1300$\,AU).

Figure~\ref{fig:2} shows that the outflow has a wide opening angle inside the host cloud, while it has good collimation outside the host cloud.
An hourglass-like configuration of field lines is realized inside the host cloud, because the field lines converge toward the center as the cloud collapses.
The gas flows along the hourglass-like lines inside the host cloud and the outflow has a wide opening angle.
On the other hand, gas flows along the straight field lines and has good collimation outside the host cloud.
The opening angle of the outflow inside the host cloud strongly influences the mass accretion rate and star formation efficiency \citep{matzner00}.

The density and velocity distribution around and inside the host cloud are plotted in Figure~\ref{fig:3}.
Figure~\ref{fig:3}{\it a} indicates that the residual matter in the host cloud is mainly distributed in a region along the rotation axis (i.e., $z$-axis) and on the equatorial ($z=0$) plane.
The source of the matter near the rotation axis is the outflow, while the matter on the $z=0$ plane corresponds to the disk which is supported by the centrifugal force. 
At the end of the calculation, $M_{\rm res}=104\mj$ of the residual mass remains inside the host cloud ($r < 2\rcri$).
Since the mass of the proto-brown dwarf is $M_{\rm ps} = 45\mj$ and the initial host cloud mass is $M_{\rm ini}=225\mj$, the mass ejected from the host cloud is $M_{\rm ej} = M_{\rm ini} - M_{\rm res}- M_{\rm ps} = 76\mj$.
In addition, since the outflowing mass is $M_{\rm out}= 128\mj$, the mass swept by the outflow into interstellar space is $M_{\rm sw} = M_{\rm out} - M_{\rm ej} = 52\mj$.
Thus, in the host cloud, 34\% ($=M_{\rm ej}/M_{\rm ini}$) of the total mass is ejected by the outflow, while 20\% ($=M_{\rm ps}/M_{\rm ini}$) is converted into the star (or proto-brown dwarf).

In addition, at the end of the calculation, a large fraction ($M_{\rm esc}=88\mj$) of the residual mass ($M_{\rm res} = 104 \mj $) has a positive flow velocity ($v_r >0$) and escapes from the host cloud.
Note that we defined the outflowing mass $M_{\rm out}$ as the gas with $v_r > c_{\rm s}$ in Figure~\ref{fig:1}, while we defined the escaping mass $M_{\rm esc}$ as the gas with $v_r > 0$ inside the host cloud.
The rest of the mass (i.e., the gas with $v_r < 0$) inside the host cloud is $16\mj$ ($= M_{\rm res} - M_{\rm esc}$), which is composed of two parts: the circum-brown dwarf disk and the accreting gas.
As denoted above, since the disk mass is $M_{\rm disk}=13\mj$, the mass of the accreting matter is $M_{\rm acc}=3\mj$.
Thus, the mass of the accreting matter is 3\% ($=M_{\rm acc}/M_{\rm res}$) of the residual mass [or 1\% ($=M_{\rm acc}/M_{\rm ini}$) of the mass of the initial host cloud].
As a result, it is considered that the gas accretion is nearly finished.
However, since the gas supply into the disk does not completely halt, (weak) outflow continues until the end of the calculation.

In Figure~\ref{fig:3}{\it a}, we divided the host cloud into three zones, according to our results.
The outflowing gas has a supersonic velocity ($v_r > c_{\rm s}$) in the outflowing zone, while it has a subsonic velocity ($0<v_r<c_{\rm s}$) in the escaping zone.
Note that almost all the matter in the escaping zone exceeds the escape velocity of the host cloud.
A part of the matter in the escaping zone comes from outflow from the disk, while the matter near the border between the inflow and outflow (i.e., near the orange contour of $v_r=0$) escapes from the host cloud by the pressure gradient force.
Since 34\% of the total mass is already ejected from the host cloud, the gravity inside the host cloud weakens.
As the initial state, we adopt a nearly equilibrium state, in which the gravity is balanced with the thermal pressure gradient force.
Thus, owing to a decrease of mass (or gravity), the matter near the envelope escapes as the outflow continues.
While the outflow remains inside the host cloud, the gas in the escaping zone in Figure~\ref{fig:3}{\it a} has a negative radial velocity ($v_r < 0$) and falls toward the proto-brown dwarf.
After the outflow penetrates the host cloud and a large fraction of mass is ejected, the radial velocity gradually decreases and finally becomes positive ($v_r > 0$), thereafter the gas escapes.

Figure~\ref{fig:3}{\it b} and \ref{fig:3}{\it c} are close-up views of Figure~\ref{fig:3}{\it a}.
As seen in Figure~\ref{fig:3}{\it b}, strong outflow appears in the region of $r < 100$\,AU.
In addition, a cavity wall appears around the border between the outflow and inflow.
During the calculation, strong outflow continues intermittently.
Figure~\ref{fig:3}{\it c} shows the roots of the outflow located at $\sim10$\,AU, far from the proto-brown dwarf.

\section{Discussion}
\begin{figure}
\includegraphics[width=150mm]{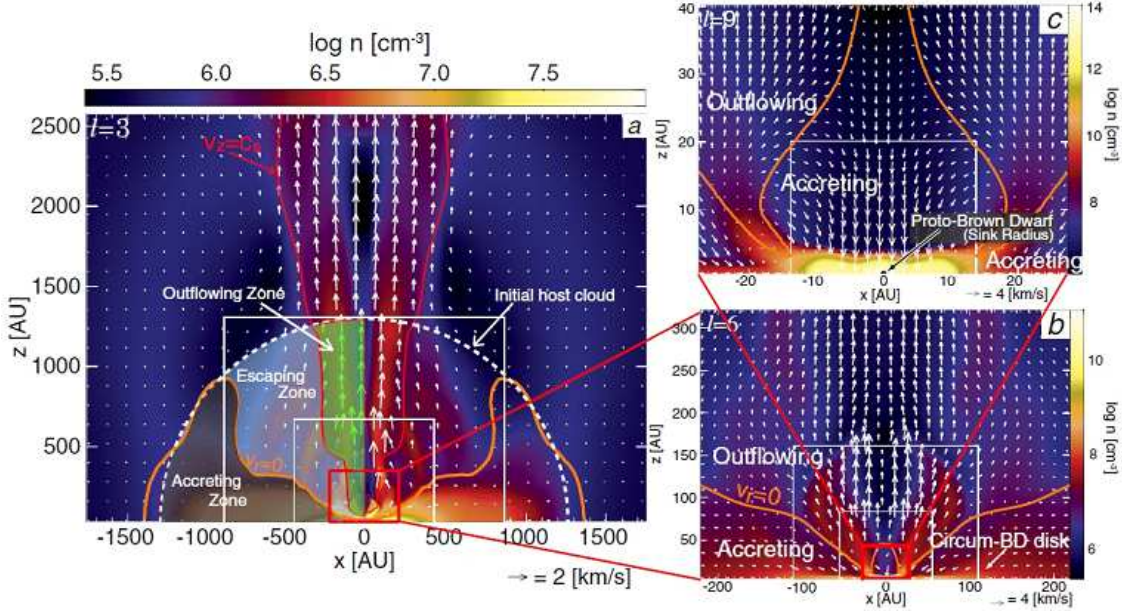}
\caption{
Density ({\it color scale}) and velocity ({\it arrows}) distribution on the $y=0$ plane at $t=12.71\,t_{\rm ff}$ ($\simeq 3.18\times 10^4$\,yr) for different grids of $l=3$ ({\it a}), 6 ({\it b}), 9 ({\it c}).
The orange contour in each panel corresponds to the boundary between the outflowing ($v_{\rm r} > 0$) and inflowing ($v_{\rm r} < 0$) gas.
The red contour in panel ({\it a}) indicates $v_z = c_{\rm s}$, inside which the gas is outflowing with supersonic speed.
The white dotted circle in panel ({\it a}) indicates the radius of the initial host cloud $r=2\rcri$.
The sphere inside $r < 2\rcri$ (i.e., inside the white dotted circle) is divided into three regions: the accreting ($v_r < 0$), escaping ($0 < v_r < c_{\rm s}$) and outflowing ($v_r > c_{\rm s}$) zones.
}
\label{fig:3}
\end{figure}

Recently, \citet{phan08} observed bipolar molecular outflow from a proto-brown dwarf ($\rho$ Oph 102) with $\sim 60\mj$.
The outflow extends up to $\sim1000$\,AU with a maximum velocity of $2.2\km$ and mass of $0.17\mj$.
In addition, they estimated that the circum-brown dwarf disk has a size of $\sim80$\,AU and mass of $8.3\mj$.
In our calculation, the outflow driven from the proto-brown dwarf with $45\mj$ has a typical flow speed of $\sim2\km$ (Fig.~\ref{fig:3}{\it b}) and mass of $128\mj$, and extends up to $\sim5\times10^3$\,AU.
The circum-brown dwarf disk in our calculation has a mass of $M_{\rm disk} = 13\mj$ and size of $\sim150$\,AU.
Thus, our results are quantitatively consistent with observations, except for the outflowing mass.

\citet{phan08} and \citet{whelan05} estimated the mass accretion rate to be $\dot{M}\sim10^{-9}\mdot$, which is comparable to our result in the later accretion phase ($\tc > 1.5\times 10^4$\,yr). 
However, their scenario for brown-dwarf formation is significantly different from our results.
We consider that a part of a brown dwarf is formed in a (small, compact) cloud core in a similar process as low-mass star formation.
However, in observational studies it seems to be considered that the brown dwarf formation is a scaled down version of that of hydrogen-burning stars, in which a very small rate of mass accretion ($\dot{M}\sim10^{-9}$--$10^{-12}\msun$) is assumed in the main accreton phase.
We show that, even in the brown dwarf formation process, the mass accretion rate remains as high as $\dot{M}\sim10^{-6}\mdot$ for $\tc \lesssim 10^4$\,yr, and it rapidly drops to $\dot{M}\sim10^{-9}$--$10^{-12}\mdot$ for $\tc \gtrsim 10^4$\,yr.

In the theoretical framework of star formation, the mass accretion rate can be described as $\dot{M}=f\, c_s^3/G $, where $f$ is a constant (e.g., $f=0.975$ for \citealt{shu77}, $f= 46.9$ for \citealt{hunter77}).
Since gas clouds have temperatures of $T\gtrsim10$\,K ($c_s\gtrsim0.2\km$), the accretion rate in the main accretion phase is $\dot{M}\gtrsim10^{-6}\mdot$ which is the minimum value attained in the general star formation scenario.
Our results show, in the main accretion phase ($\tc < 1.5 \times 10^4$\,yr), the accretion rate of $\sim2\times 10^{-5}-10^{-6}\mdot$\ that corresponds to $f\simeq1-10$.
Note that other effects such as cloud rotation, magnetic field, and turbulence only increase the mass accretion rate.
Thus, the scenario in which the accretion rate of $10^{-9}\mdot$ lasts for $\sim10^7$\,yr is not theoretically expected. 
On the contrary, it is reasonable that an accretion rate of $\sim10^{-6}\mdot$ lasts for $\sim10^4$\,yr for brown-dwarf formation.
In this case, since the main accretion phase is very short, it is difficult to observe the accretion onto a proto-brown dwarf with $\dot{M} \sim 10^{-6}\mdot$.
Thus, we may often observe low accretion rates of $<10^{-9}\mdot$, because a lower accretion phase is expected to last for a longer duration ($\gg 10^4$\,yr).
In this study, the accretion rate drops at $\tc \gtrsim 10^4$\,yr because we adopted a less massive core as the initial state.
However, we expect that if a massive core is adopted as the initial state, a high accretion rate of $\sim 10^{-6} \mdot$ lasts for a long duration of $t_{\rm c}\simeq 10^5$--$10^6$\,yr, and a stellar mass star forms.
The result in this letter favors the idea that the stellar mass is determined by the initial size of the host cloud and hence brown dwarf sized objects are formed in less massive clouds.

\begin{figure}
\includegraphics[width=150mm]{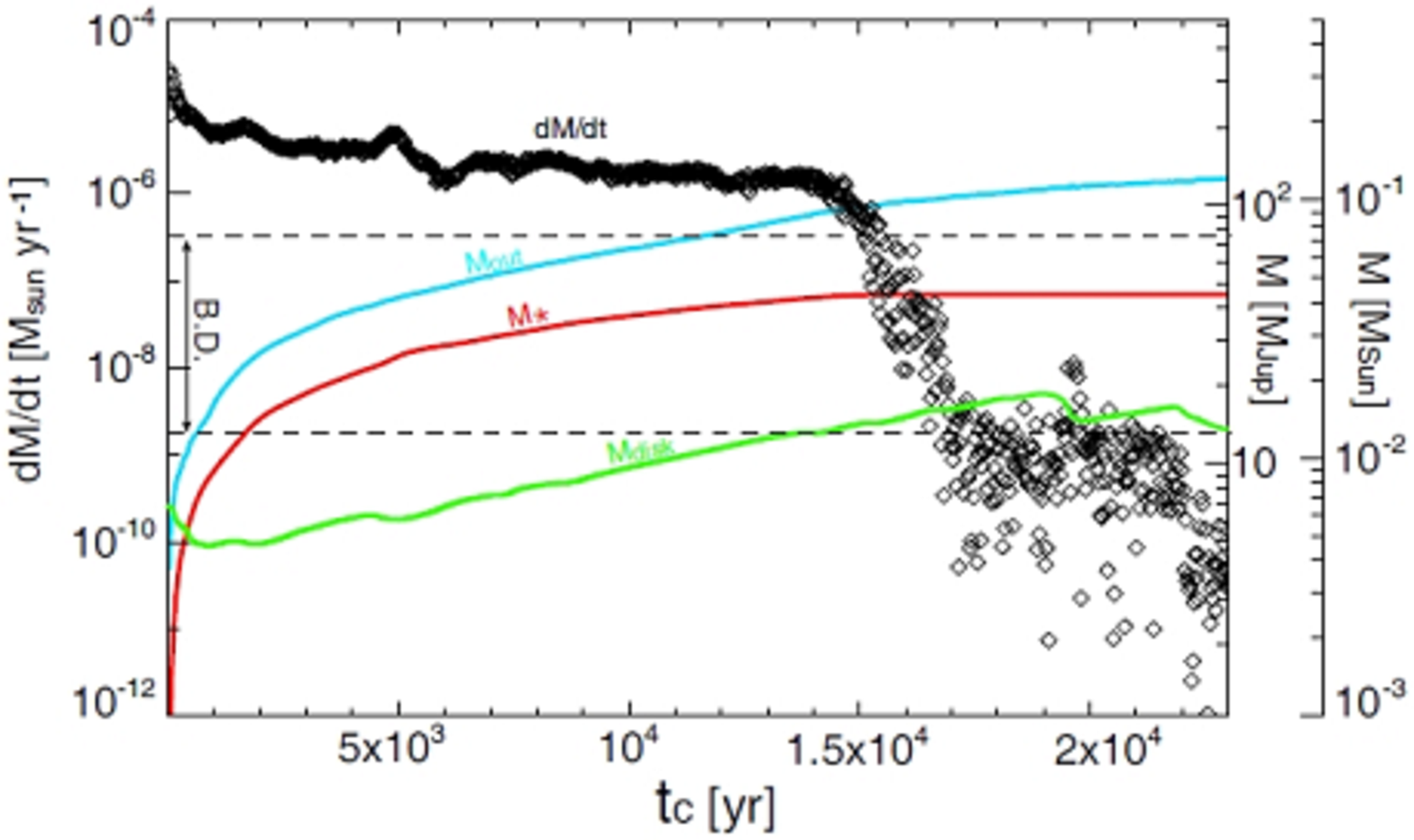}
\caption{
Mass accretion rate ({\it diamonds}), and mass of the proto-brown dwarf ({\it red}), circum-brown dwarf disk ({\it green}), and outflowing gas ({\it blue}) against the elapsed time after the proto-brown dwarf formation.
The left axis indicates the mass accretion rate, while the right axes indicates mass in units of Jovian and solar masses.
The two horizontal dashed lines indicate the lower and upper limits of brown dwarf mass.
}
\label{fig:1}
\end{figure}

We comment on the sudden drop of the mass accretion rate at $\tc\sim10^4$\,yr, as seen in Figure~\ref{fig:1}.
The epoch of the sudden drop almost corresponds to the epoch at which the outflow escapes from the host cloud.
Since the outflow with a significant mass is escaped from the host cloud by this epoch, the gravitational potential of the host cloud 
shallows and the accretion rate weakens, which causes the sudden drop of the accretion rate.
We expect that the epoch of this sudden drop depends on the size of the host cloud.

For the outflow model, \citet{phan08} proposed a jet-driven bow shock model \citep{masson93}.
However, our results are different from the predictions of this model.
Our calculations show that the outflow is directly driven from the circum-brown dwarf disk.
No appearance of the high-speed jet in our calculation is due to the fact that we adopted the region of $r<0.5$\,AU as the sink, and thus, we did not calculate the region near the proto-brown dwarf.
Adopting a central object with a mass of $40\mj$, the Kepler speed at $\sim0.5$\,AU is $\sim10\km$.
Thus, flow of $>10\km$ could not be resolved in our calculation.
In contrast, \citet{machida08b} calculated the structure in the near proximity of the protostar and showed that a high-speed jet of $\sim30\km$ is driven from the Jovian-mass protostar.
Thus, when we resolve the structure in the near proximity to the proto-brown dwarf, a high-speed jet is expected to be driven from the proto-brown dwarf.
As shown in \citet{machida08b}, we expect that the high-speed jet hardly affects the mass accretion rate and star formation efficiency, because it has a well-collimated structure.
Our calculations show that the wide-opening molecular outflow driven from the circum-brown dwarf disk strongly affects the mass accretion rate and star formation efficiency, as predicted in \citet{matzner00}.

\acknowledgments
This work was supported by a Grant-in-Aid for the Global COE Program ``The Next Generation of Physics, Spun from Universality and Emergence'' from the Ministry of Education, Culture, Sports, Science and Technology (MEXT) of Japan, and partially supported by Grants-in-Aid from MEXT (18740104, 20540238, 21740136).

\clearpage

\end{document}